\documentclass[11pt]{article}

\usepackage{moriond,epsfig}
\usepackage{amsmath,amssymb}
\usepackage{subfigure}

\def\ab{\bar{\alpha}}
\def\vx{\mathbf{x}}
\def\vy{\mathbf{y}}
\def\vz{\mathbf{z}}
\def\vr{\mathbf{r}}
\def\vk{\mathbf{k}}
\def\vq{\mathbf{q}}
\def\vp{\mathbf{p}}

\newcommand{\abs}[1]{\left|#1\right|}
\def\N{{\cal N}}
\def\tN{\tilde{\cal N}}

\begin{document}
\vspace*{4cm}
\title{Geometric scaling in high-energy QCD at nonzero momentum transfer}

\author{G. Soyez\footnote{On leave from the fundamental theoretical physics group of the University of Li\`ege.}, C. Marquet, R. Peschanski}

\address{SPhT, CEA/Saclay, Orme des Merisiers, Bat 774, F-91191 Gif-sur-Yvette cedex, France}

\maketitle
\abstracts{
We show how one can obtain geometric scaling properties from the Balitsky-Kovchegov (BK) equation. We start by explaining how, this property arises for the $b$-independent BK equation. We show that it is possible to extend this model to the full BK equation including momentum transfer. The saturation scale behaves like max$(q,Q_T)$ where $q$ is the momentum transfer and $Q_T$ a typical scale of the target.
}

Geometric scaling \cite{geom} expresses the fact that the DIS HERA data depend of the virtuality $Q^2$ and the rapidity $Y=\log(1/x)$ only through the ratio $Q/Q_s(Y)$ where $Q_s$ is an energy-dependent scale.
On another side, a lot of research have been done during the last ten years to obtain a description of the saturation effects in the high-energy limit of QCD, leading to the Balitsky-Kovchegov equation \cite{bk}. This equation supplements the BFKL \cite{bfkl} equation in including the contributions from fan diagrams. It has recently been shown in the framework of the impact-parameter independent BK equation that one can infer geometric scaling from the asymptotic solutions of the evolution equation. 

We shall explain how we have extended these properties of geometric scaling to the case of the full BK equation. Our analysis is performed in momentum space where we also introduce a new form of the BK equation. We shall discuss the properties of the saturation scale and its momentum-transfer dependence.

\begin{figure}
\subfigure[$\log(Rq)=-5$]{\includegraphics[scale=0.75]{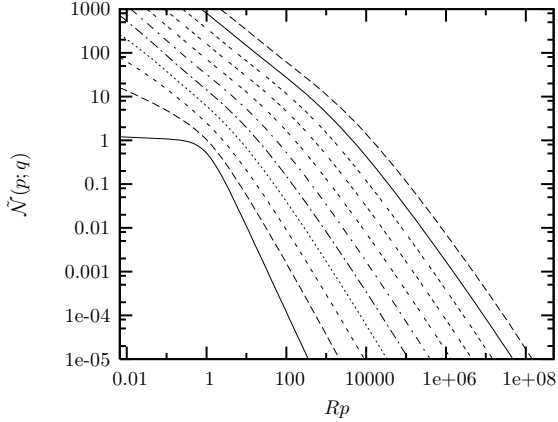}}\hspace{1.5cm}
\subfigure[$\log(Rq)=2$]{\includegraphics[scale=0.75]{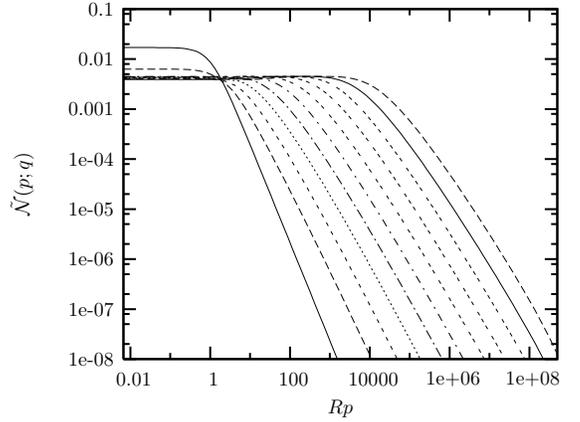}}
\caption{Rapidity evolution of the dipole density as a function of $p$ for different values of $q$ (see text). For each plot, we show the amplitude for $Y$ varying between 0 and 25 by steps of 2.5.}
\label{fig:front}
\end{figure}

Let us thus start our analysis by considering the case of the impact-parameter independent BK equation: $\N(\vx,\vy) = \N(\abs{\vx\!-\!\vy})$. If we go to momentum space using
\[
\tN(k) = \frac{1}{2\pi}\int \frac{d^2r}{r^2}\,e^{i\vk.\vr}\N(r)
       = \int_0^\infty \frac{dr}{r}\,J_0(kr)\N(r),
\]
the BK equation takes the form \cite{us2}
\[
\partial_Y \tN(k) = \ab\chi(-\partial_L)\tN(k) - \ab \tN^2(k),
\]
where $\chi(\gamma)=2\psi(1)-\psi(\gamma)-\psi(1-\gamma)$ is the BFKL kernel and $L=\log(k^2/k_0^2)$. If we expand this kernel to second order around $\gamma=1/2$, this equation becomes \cite{mp} equivalent to the Fisher-Kolmogorov-Petrovsky-Piscounov equation \cite{fkpp}. This equation, well-known is statistical physics, admits traveling-wave solutions at large rapidities {\em i.e.} the solution depends only on $Y-v_c\log(k)$, where $v_c$ is a constant. More generally, if the evolution equation satisfies the following conditions:
\begin{itemize}
\item $\tN=0$ is an unstable fix point and the equation contains a linear term, leading to a growth of the amplitude, and a non-linear damping term;
\item the initial condition is steep enough, which is ensured by colour transparency,
\item the linearised equation admits a superposition of waves as solution 
\begin{equation}\label{eq:waves}
\tN_{\text{lin}}(k,Y) = \int_{c-i\infty}^{c+i\infty} \frac{d\gamma}{2i\pi} a_0(\gamma) \exp\left[\omega(\gamma) Y - \gamma \log(k^2/k_0^2)\right],
\end{equation}
\end{itemize}
then the asymptotic solution of the full equation takes the form
\begin{equation}\label{eq:travwaves}
\tN(k,Y) \underset{Y\to\infty}{\sim} \left[\frac{k^2}{Q_s^2(Y)}\right]^{-\gamma_c},
\end{equation}
for $k\gg k_0$ where the saturation scale $Q_s^2(Y)$ grows as $k_0\ e^{v_c Y}$. The critical exponent $\gamma_c$ and the critical speed $v_c$ are given by $v_c = \omega(\gamma_c)/\gamma_c = \omega'(\gamma_c)$, meaning that the group velocity is equal to the phase velocity.

\begin{figure}
\includegraphics[scale=0.85]{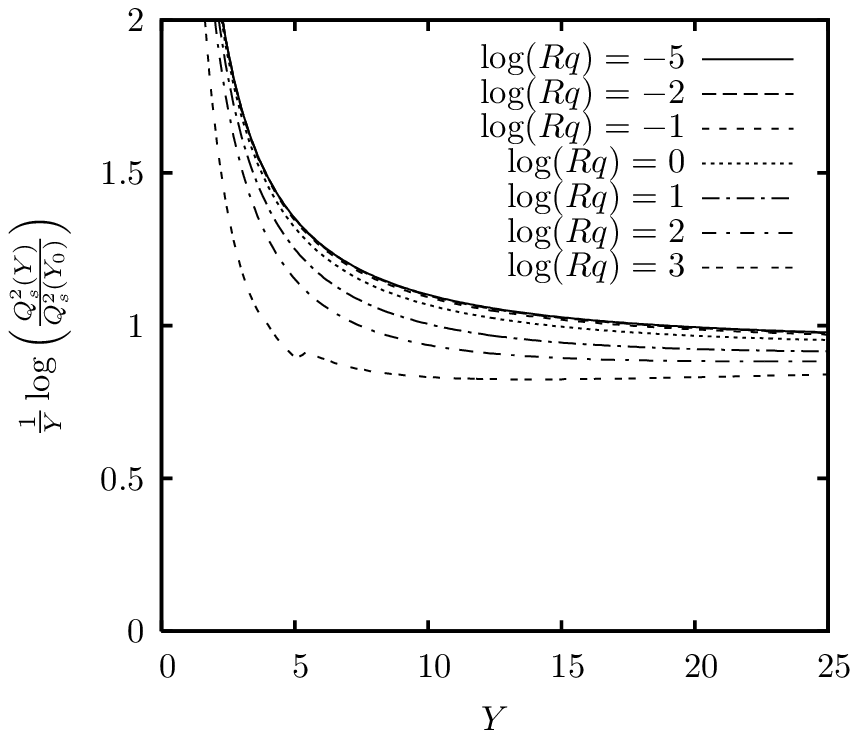}
\includegraphics[scale=0.85]{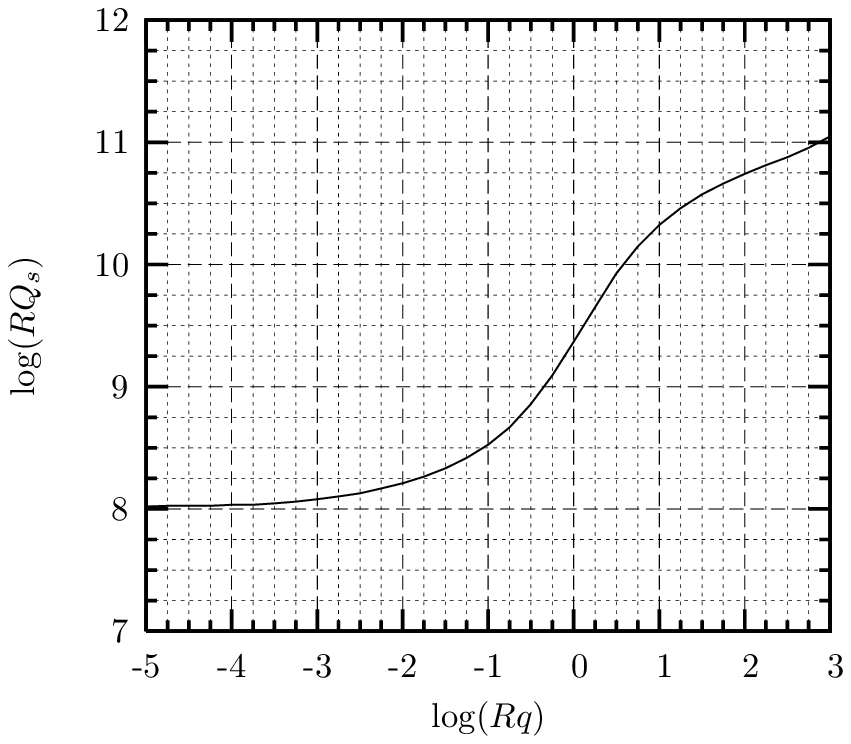}
\caption{Properties of the saturation scale. Left: evolution of the saturation scale with $Y$. Right: $q$-evolution of the saturation scale (see text).}
\label{fig:qs}
\end{figure}

The aim of our studies \cite{us,us2} is to show that the properties of having traveling waves as asymptotic solutions holds for the complete BK equation
\[
\partial_Y \N(\vx,\vy) = \frac{\ab}{\pi}\int d^2f\, \frac{(\vx-\vy)^2}{(\vx-\vz)^2(\vz-\vy)^2}
\left[\N(\vx,\vz)+\N(\vz,\vy)-\N(\vx,\vy)-\N(\vx,\vz)\N(\vz,\vy) \right].
\]
It directly appears that, already at the level of the linear terms, this equation is non-local in impact parameter. Our first step is thus to move to momentum space
\[
\tN(\vk,\vq) = \frac{1}{(2\pi)^2}\int d^2x\, d^2y\,e^{i\vk.\vx}e^{i(\vq-\vk).\vy}
  \frac{\N(\vx,\vy)}{(\vx-\vy)^2}\,.
\]
After some small manipulations, the equation takes the form
\begin{eqnarray}\label{eq:bkmom}
\partial_Y \tN(\vk,\vq)
 & = &\frac{\bar\alpha}{\pi}\int \frac{d^2k'}{(\vk-\vk')^2}
   \left\{
   \tN(\vk',\vq) - \frac{1}{2} \left[
     \frac{\vk^2}{\vk'^2+(\vk-\vk')^2} +
     \frac{(\vq-\vk)^2}{(\vq-\vk')^2+(\vk-\vk')^2}
     \right] \tN(\vk, \vq)
   \right\} \nonumber \\
 & - & \frac{\bar\alpha}{2\pi} \int 
d^2k'\,\tN(\vk,\vk')\,\tN(\vk-\vk',\vq-\vk').
\label{eq:bk}
\end{eqnarray}
Due to the fact that in this equation the effect of the linear kernel remains local in $q$, to the contrary of what happens in coordinate space, the properties of geometric scaling and the formation of traveling waves is better studied in momentum space. 

We now need to check that the three conditions introduced in the previous sections are indeed fulfilled in the case of the complete BK equation. The only technical point is obviously to check that the solution of the linear equation can be written under the form \eqref{eq:waves} of a superposition of waves. Actually, one can show that the solutions of equation \eqref{eq:bkmom}, reduced to the linear terms, can be put into the form
\begin{equation}\label{eq:sollin}
\tN_{\text{lin}}(\vk,\vq)=\int_{\frac{1}{2}-i\infty}^{\frac{1}{2}+i\infty} \frac{d\gamma}{2i\pi}\ e^{\bar\alpha\chi(\gamma)Y}
f^\gamma(\vk, \vq)\phi_0(\gamma,\vq),
\end{equation}
where $f^\gamma(\vk, \vq)$ is known analytically and $\phi_0(\gamma,\vq)$ contains all reference to the initial condition. Although $f^\gamma(\vk,\vq)$ is a rather complicated function, in the limit $\abs{k}\gg\abs{q}$ it simply becomes proportional to $(k/q)^{-2\gamma}$. This means that in the latter limit, the condition \eqref{eq:waves} is satisfied with $k_0 = q$. Hence, when rapidity increases, traveling waves get formed with the same critical properties as in the $b$-independent case. The saturation scale is
\[
Q_s^2(Y) = q^2 \Omega_s^2(Y) \underset{Y\to\infty}{\sim} q^2
\exp\left[\bar\alpha\frac{\chi(\gamma_c)}{\gamma_c}Y-\frac{3}{2\gamma_c}\log(Y)
-\frac{3}{\gamma_c^2}\sqrt{\frac{2\pi}{\bar\alpha \chi''(\gamma_c)}} \frac{1}{\sqrt{Y}}
+{\cal{O}}(1/Y)\right]
\]
and the tail of the wavefront is then given by
\[
\tN(\vk,\vq) \underset{Y\to\infty}{\sim}
\phi^{\gamma_c}(\vq)\,\log\left(\frac{k^2}{q^2\Omega_s^2(Y)}
\right)\,\abs{\frac{k^2}{q^2\Omega_s^2(Y)}}^{-\gamma_c}\,
\exp\left[-\frac{1}{2\bar\alpha\chi''(\gamma_c)Y}\log^2\left(
\frac{k^2}{q^2\Omega_s^2(Y)}\right)\right].
\]

Due to the presence of the $\phi_0(\gamma,\vq)$ in \eqref{eq:sollin}, when $q$ becomes smaller than the characteristic size $Q_T$ of the target, one gets an extra-factor $\abs{Q_T/q}^{2\gamma}$. As a consequence, $Q_T$ substitutes to $q$ as the reference scale $k_0$. The saturation scale $Q_s$ becomes is therefore proportional to $q$ or $Q_T$ whether $q$ is larger or smaller than $Q_T$.

We have performed \cite{us2} numerical simulations of the full BK equation in momentum space \eqref{eq:bkmom}. For the purposes of these studies, we have introduced $R = Q_T^{-1}$, the typical size of the target and $\vp=\vk-\vq/2$, canonically conjugated to the dipole size $\vr$. The strong coupling constant has been fixed to $\ab=0.2$. As clearly shown in figure \ref{fig:front}, we have formation of a traveling wave when $k\gg q$ and at large rapidities. To observe this fact more precisely, we have constructed the saturation scale $Q_s$ and studied its rapidity evolution. The result is presented in figure \ref{fig:qs} and shows, as expected, goes to $v_c=4.886 \ab=0.9772$. 

Finally, figure \ref{fig:qs} also shows the $q$ dependence of the saturation scale at large rapidity ($Y=25$). The curve shows that $Q_s$ goes to a constant value when $q\ll Q_T$ and then takes a linear behaviour. The very-large-$q$ part of this curve deviates from this behaviour due to the fact that the asymptotic regime is not yet reached.

To summarize, we have been able to obtain traveling waves solutions for the full BK equation. For locality reasons, our analysis is done in momentum space. We predict geometric scaling in terms of $k/(q\Omega_s(Y))$ or $k/(Q_T\Omega_s(Y))$ whether $q$ is larger or smaller than $Q_T$. The rapidity dependence of the saturation scale remains the same as in the $b$-independent situation.

To conclude, numerical simulations would allow for testing these observations phenomenologically. On theoretical grounds, one can include additional effects such as the running of the coupling, the odderon effects or the fluctuation term.

\section*{Acknowledgements}
G.S. is funded by the National Funds for Scientific Research (Belgium).

\end{document}